# Prospects of Hysteresis-Free Abrupt Switching ($0mV/dec$) in Landau Switches

Ankit Jain and Muhammad Ashraful Alam[*]
School of ECE, Purdue University, West Lafayette, IN, USA, 47907
e-mail[*]:alam@purdue.edu

*Abstract*-**Sub-threshold swing ($S$) defines the sharpness of ON-OFF switching of a Field Effect Transistor (FET) with $S = 0$ corresponding to abrupt switching characteristics. While thermodynamics dictates $S \geq 60mV/dec$ for classical FETs, "Landau switches" use inherently unstable gate insulators to achieve abrupt switching. Unfortunately, $S = 0$ switching is always achieved at the expense of an intrinsic hysteresis, making these switches unsuitable for low-power applications. The fundamental question therefore remains: Under what condition, *hysteresis-free abrupt switching* can be achieved in a Landau switch? In this paper, we first provide an intuitive classification of all charge based switches in terms of their energy landscapes and identify two-well energy landscape as the characteristic feature of Landau switches. We then use nanoelectromechanical field effect transistor (NEMFET) as an illustrative example of a Landau switch and conclude that a flat energy landscape is essential for hysteresis-free abrupt switching. In contrast, a hysteresis-free *smooth* switching ($S \leq 60mV/dec$) is obtained by stabilizing the unstable gate insulator in its unstable regime, so as to provide internal voltage amplification necessary to achieve $S \leq 60mV/dec$. Our conclusions have broad implications and may considerably simplify the design of next charge based logic switch.**

## I. INTRODUCTION

Scaling of field effect transistors (FETs) may eventually come to an end due to the non-scalability of sub-threshold swing ($S$); whose minimum value is thermodynamically limited to $60mV/dec$ at room temperature. There is a worldwide search for the next logic switch [1] that can overcome the "Boltzmann tyranny" [2] of classical FETs and reduce $S$ below the Boltzmann limit. Sub-threshold swing ($S$) of any FET is defined as the change in gate voltage ($V_G$) required for one order change in the drain current ($I_{DS}$), i.e.

$$S \equiv \frac{dV_G}{d\log_{10}(I_{DS})} = \left(\frac{dV_G}{d\psi_s}\right)\left(\frac{d\psi_s}{d\log_{10}(I_{DS})}\right) = m \times n, \quad (1)$$

where transport factor $n \equiv \frac{d\psi_s}{d\log_{10}(I_{DS})}$ is dictated by the current transport in the channel and body factor $m \equiv \frac{dV_G}{d\psi_s}$ governs the coupling between the gate voltage ($V_G$) and channel potential ($\psi_s$). The value of $m$ can be obtained by the schematic and equivalent capacitor divider model shown in Figs. 1A-B, and is given by-

$$m \equiv \frac{dV_G}{d\psi_s} = 1 + \frac{C_s}{C}, \quad (2)$$

where $C_s$ is the depletion capacitance, $C \equiv C_{ox}C_{ins}/(C_{ox} + C_{ins})$ is the series capacitance of gate insulator ($C_{ins} = \epsilon/y$) and thin $SiO_2$ layer ($C_{ox}$). Here, $\epsilon$ and $y$ are respectively the permittivity and thickness of the gate insulator.

In classical Boltzmann switch, top of the barrier transport in the channel dictates $n = 60mV/dec$ [3] at room temperature and traditional gate insulator with $C_{ins} > 0$ dictates $m > 1$ (Eq. 2). These two facts combined limit $S \geq 60mV/dec$ for classical FETs (Eq. 1). In literature, there have been two major approaches to reduce $S$ below $60mV/dec$ (Fig. 1C). The first scheme involves reducing $n < 60 \ mV/dec$ (while keeping $m$ fixed) by modifying the transport within the channel (e.g. source to channel tunneling in tunnel-FETs [4], impact ionization in I-MOS [5], etc.). In the second approach, $m$ is reduced below one by changing the gate-insulator. In this scheme, the classical gate insulator ($C_{ins}$) characterized by a single well energy landscape as shown in Fig. 1D is replaced by an inherently unstable gate insulator which exhibits a two-well energy landscape, Fig. 1E. *We call these new class of switches Landau switches, because of the similar two well energy landscape associated with phase transition* [6]. Known examples of Landau switches include Ferroelectric-FET (FE-FET) [7], and nanoelectromechanical field effect transistor (NEMFET) [8].

Gate insulator is a ferroelectric material in FE-FET whereas NEMFET has an air-gap as the gate insulator is (Fig. 2). In FE-FET, instability in $C_{ins} = \epsilon/y$ arises due to instability in $\epsilon$, whereas in NEMFET due to instability in $y$, due to pull-in instability [9]. The two wells ($W_1 \& W_2$) refer to two polarization states ($P$) in the ferroelectric, whereas they refer to gate up ($y = y_0$) and down ($y = 0$) positions in NEMFET. The abrupt switching from one well (e.g. $W_1$) to another (e.g. $W_2$) causes $m = 0$ (the role of inherent instability) and thus gives rise to abrupt switching characteristics equivalent to $0mV/dec$ as shown in Fig. 1F.

In principle, abrupt switching of Landau switches could potentially reduce the energy dissipation to the absolute minimum (if operated between points $O_1$ and $O_2$ in Fig. 1F). But, abrupt switching always comes at the cost of an intrinsic hysteresis, because once switched from $W_1$ to $W_2$ (or vice versa), switching back from $W_2$ to $W_1$ does not occur at the same applied voltage. Therefore, hysteresis in Landau switches (Fig. 1F) dictates the minimum energy dissipation ($E_d = Q_V H_V$ with $Q_V$ being the difference in charge between two states and $H_V$ being the width of the hysteresis), because the switch must operate between points $O_3$ and $O_2$ (Fig. 1F). *The fundamental question therefore remains under what condition hysteresis-free abrupt switching ($H_V = 0, S = 0mV/dec$) can be realized in a Landau switch leading to minimum energy dissipation (when operated between points $O_1$ and $O_2$ in Fig. 1G)?*

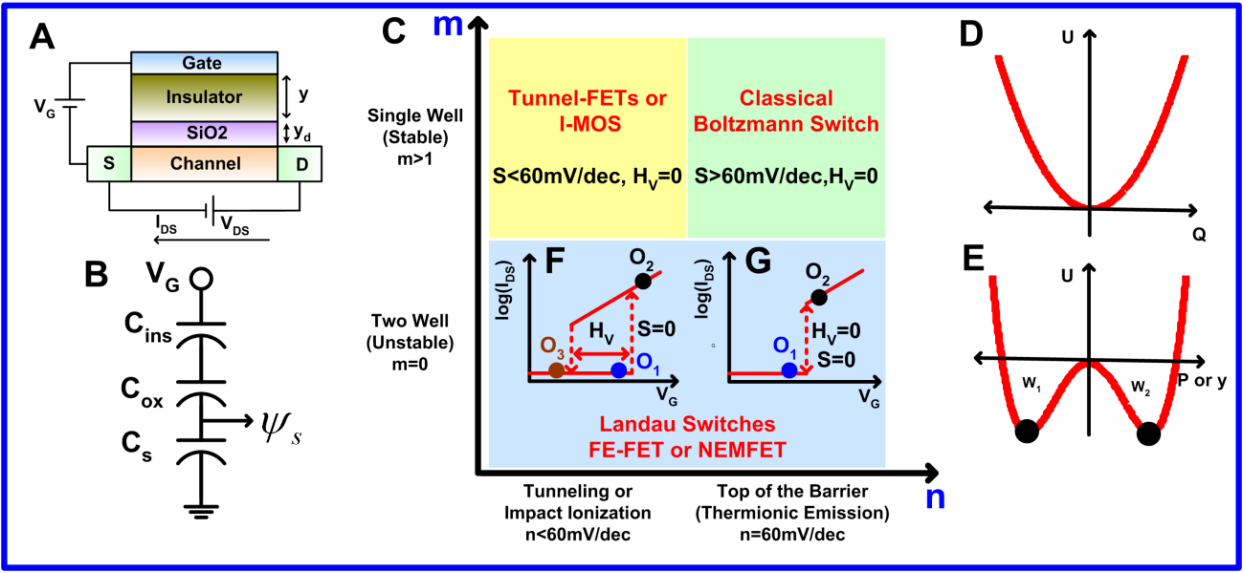

**Fig. 1:** Phase space of FETs. **(A)** Schematic of a generic FET and **(B)** its equivalent capacitor divider model. **(C)** Classification of various proposals of novel field effect transistors based on the values of $m$ and $n$. $H_V$ is the width of the hysteresis in $I_{DS} - V_G$ characteristic. FETs with stable gate insulator ($C_{ins}$) having single well energy landscape **(D)** exhibits $m > 1$. $U$ is the total energy and $Q$ is the charge on $C_{ins}$. FETs with an inherently unstable gate insulator having two well energy landscape **(E)**, exhibits $m = 0$ during switching from one well to another well. $P$ is the polarization of Ferroelectric material and $y$ is the position of gate in NEMFET. Inherent instability of gate insulator gives rise to abrupt switching characteristics **(F)**, however, always at the cost of hysteresis limiting energy dissipation to $E_d = Q_V H_V$. The fundamental question is: Under what conditions hysteresis-free abrupt switching can be achieved in Landau switches? **(G)**.

We consider this question in a general context, because the energy landscape of all Landau switches is characterized by two isolated energy wells separated by an energy barrier (Fig. 1E) and therefore the conclusions reached by analyzing one switch (e.g. NEMFET) should apply broadly to all other switches of this class. We emphasize that abrupt switching in Landau switches occur irrespective of the transport mechanism within the channel. In this article, therefore, we restrict ourselves to $n = 60mV/dec$ of a typical Boltzmann switch to focus exclusively on strategies that make $m < 1$. The paper is organized as follows. In section II, we discuss the physics of hysteresis in NEMFET. The question of hysteresis-free $0mV/dec$ switching is addressed in section III. Section IV shows a unique material independent way of achieving hysteresis-free sub $60mV/dec$ switching in NEMFET. In section V, we explore the prospects of hysteresis-free abrupt switching in FE-FET to verify whether the conclusions based on NEMFET are general enough to be applicable to another example of a Landau switch. Our conclusions are summarized in section VI.

## II. THEORY OF HYSTERESIS IN NEMFET

Figure 2A shows the schematic of NEMFET in which gate is a fixed-fixed beam and air-gap creates the gate insulator; the structure is similar to a suspended-gate FET [10] and resonant gate transistor [9]. The essential operation of NEMFET can be understood in terms of a spring-mass system [9] in which the movable gate is suspended from a spring as shown in Fig. 2B. When a dc bias $V_G$ is applied, the position of the gate can be obtained by minimizing the total system energy ($U = U_s + U_e$). Mechanical spring energy ($U_s$) and the electrostatic energy ($U_e$) are given by-

$$U_s = \frac{1}{2}k(y_0 - y)^2, \qquad (3)$$

$$U_e = -\frac{1}{2}\frac{\epsilon_0 A}{y + y_d^{eff}}V_G^2, \qquad (4)$$

where $k = \alpha E L H^3 / W^3$ is stiffness of the gate, $\alpha$ is a geometrical constant, $E$ is the Young's modulus of gate material, $L$ is the channel length (also equal to the gate width), $W$ is the channel width (also equal to gate length), $H$ is the thickness of the gate, $y_0$ is the initial air-gap, $y$ is the gap between the gate and $SiO_2$, $\epsilon_0$ is the permittivity of free space, $A = WL$ is area of the gate, and $y_d^{eff} = \epsilon_0/C_s$ is the effective depletion width (normalized to air) of semiconductor channel.

To understand the operation of NEMFET, let us consider the evolution of $U - y$ landscape as a function of $V_G$. For a given $V_G$, the system is stabilized at the minimum of $U$, shown by open circles in Fig. 3A. With increasing $V_G$, gate is stabilized at smaller gap and $y$ decreases (Figs. 3 A & C). When $V_G$ exceeds the pull-in voltage ($V_{PI}$), $U - y$ landscape does not exhibit any local minima making NEMFET inherently unstable. Therefore, gate can no longer be stabilized in air. The gate is now pulled-in to stabilize at $y =$ 0, corresponding to the global minima of $U$ (Figs. 3A & C). This well-known pull-in instability [9], [11] of NEMFET occurs at a critical air-gap $y_c$ given by (see section A1 of Appendix for derivation)-

$$y_c = \frac{2}{3}y_0 - \frac{1}{3}y_d^{eff}. \quad (5)$$

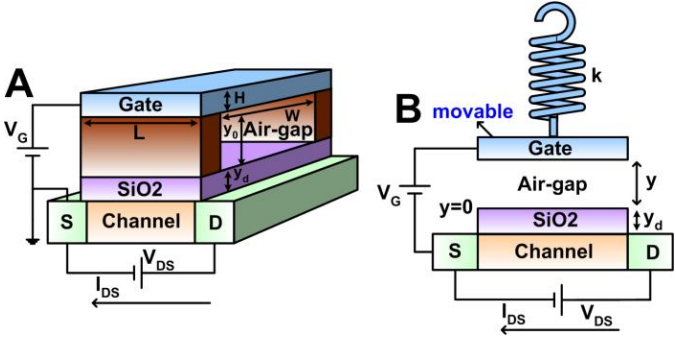

**Fig. 2:** **(A)** Schematic of NEMFET in which gate is a fixed-fixed beam suspended above the channel and **(B)** its equivalent spring-mass model in which gate is suspended from a spring of stiffness $k$. Initial gap between the gate and $SiO_2$ is $y_0$.

Note that, for $y_d^{eff} \ll 2y_0$, $y_c$ reduces to $\frac{2}{3}y_0$. In other words, gate can only be stabilized in the region $\frac{2}{3}y_0 < y < y_0$, with rest of the gap positions inaccessible due to stability considerations. This well-known result is shown schematically in Fig. 3D. The position of the gate remains clamped at $y = 0$ with further increase in $V_G$ (Fig. 3C). We stress that this discontinuous jump in $y - V_G$ characteristic (i.e., $y_c \neq 0$) makes $C_{ins} = C_{air} = \epsilon_0/y$ discontinuous, with a corresponding discontinuous jump in the $I_{DS} - V_G$ characteristic corresponding to $S = 0mV/dec$ switching characteristics.

When $V_G$ is reduced, gate does not immediately spring back in air at $V_G = V_{PI}$. This is because of the presence of an energy barrier (Fig. 3B, green curve). Therefore, gate remains at $y = 0$ (shown by open triangles in Figs. 3B-C) until $V_G$ is reduced such that energy barrier vanishes (Fig. 3B, magenta curve). This occurs at pull-out voltage ($V_{PO}$) and any reduction in $V_G < V_{PO}$, releases the gate (Figs. 3B-C). This asymmetry between pull-in and pull-out due to the presence of an energy barrier at pull-in, results in hysteretic $y - V_G$ characteristic i.e., $H_V = V_{PI} - V_{PO} \neq 0$ (Fig. 3B) and also hysteretic $I_{DS} - V_G$ characteristic as shown in Fig. 1F. Despite the $S = 0mV/dec$ transition, this $H_V \neq 0$ makes Landau switches unsuitable for ultra-low voltage applications. Given this background, we now explore the prospects of hysteresis-free ($H_V = 0$) abrupt switching $S = 0$ in Landau switches.

## III. HYSTERESIS-FREE ABRUPT SWITCHING

Hysteresis-free abrupt switching ($H_V = 0$ & $S = 0mV/dec$, see Fig. 1E) in NEMFET must display a $y - V_G$ characteristic similar to that of Fig. 4A, i.e., $H_V = 0$ and $y_c \neq 0$. Such a $y - V_G$ characteristic can be obtained only if the total energy of the system ($U_T$) is flat at the pull-in point, as shown in Fig.

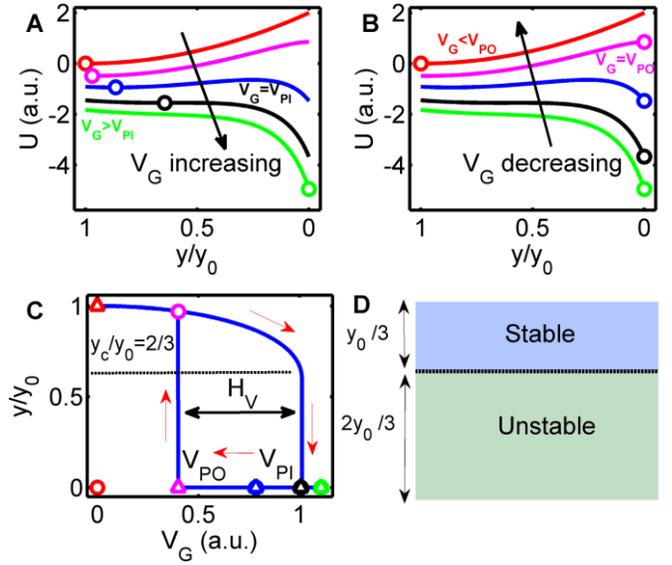

**Fig. 3:** Hysteresis and Pull-in instability in NEMFET. Energy landscapes when **(A)** $V_G$ is increasing and **(B)** $V_G$ is decreasing. Open circle and open triangle denotes the equilibrium position of the gate. **(C)** The corresponding $y - V_G$ characteristic of NEMFET showing abrupt jump due to pull-in instability, and hysteresis due to asymmetry between pull-in and pull-out. **(D)** Due to pull-in instability, gate can only be stabilized in certain region of the gap; rest of the region is inherently unstable.

4B. A flat energy landscape allows switching of the gate between $y = y_c$ and $y = 0$ with infinitesimally small change in $V_G$. Figure 4C shows (based on Eqs. 3-4) that energy landscape of a NEMFET at pull-in exhibiting energy barrier. Therefore, the requisite for flat energy landscape in Fig. 4B can only be achieved by incorporating an external energy component ($U_{ext}$) to compensate the energy barrier at the pull-in point. This additional energy can be given by-

$$U_{ext} = \begin{array}{ll} 0 & ; \ y_c < y < y_0 \\ U_1 & ; \ 0 < y < y_c \end{array}, \quad (6a)$$

$$U_1 = \frac{1}{2}\epsilon_0 A V_{PI}^2 \left( \frac{1}{y_d^{eff} + y} - \frac{1}{y_d^{eff} + y_c} \right) + \frac{1}{2}k((y_0 - y_c)^2 - (y_0 - y)^2), (6b)$$

and is plotted in Fig. 4D. Note that, $U_{ext}$ does not depend on $V_G$. In principle, $U_{ext}$ could be provided by a nonlinear spring [12] i.e., $U_s^{NL} = U_s + U_{ext}$. Indeed, an equivalent of such "nonlinear spring" to flatten the energy landscape is the pre-requisite for all Landau switches to achieve hysteresis-free abrupt switching. In NEMFET, however, it is well-known that a realistic spring (made up of linear elastic material) can only provide nonlinearity of up to third order i.e., $U_s^{real} = \frac{1}{2}k(y_0 - y)^2 + \frac{1}{4}k'(y_0 - y)^4$, where $k'$ is a geometrical constant associated with spring of cubic nonlinearity [13], [14], and therefore cannot adequately compensate the energy barrier of a typical spring-mass system (Eqs. 3-4 & 6). *Lack of such highly nonlinear spring makes us to conclude that hysteresis-free abrupt switching is practically impossible in present state-of-the-art NEMFET.* It should however inspire new material or spring designs that can compensate highly

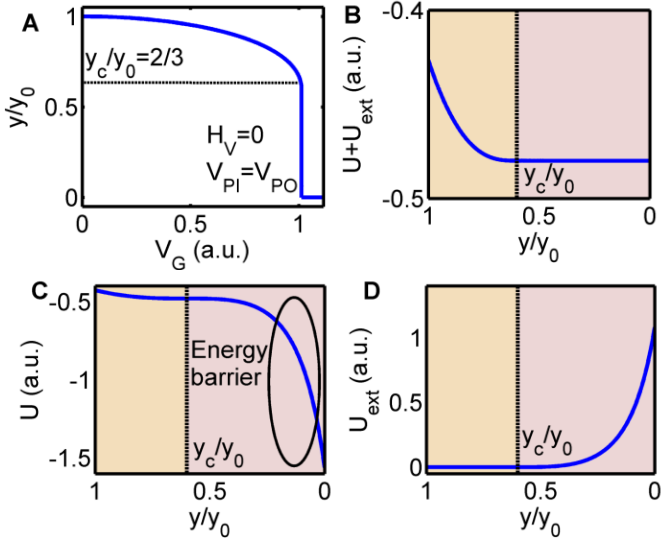

**Fig. 4:** Practical impossibility of hysteresis-free abrupt switching ($S = 0 mV/dec$) in NEMFET. **(A)** Equivalent $y - V_G$ characteristic for hysteresis-free abrupt switching and **(B)** required flat potential landscape at pull-in. **(C)** Potential landscape of regular NEMFET at pull-in showing the energy barrier which is the main cause of hysteresis and **(D)** extra potential which is required to compensate the energy barrier in (C).

nonlinear energy landscape of NEMFET to achieve hysteresis-free abrupt switching.

## IV. Hysteresis-Free Smooth Switching

In contrast to Hysteresis-free abrupt switching ($H_V = 0$ & $S = 0$), hysteresis-free smooth switching ($H_V = 0$ & $S > 0 mV/dec$) requires a benign $y - V_G$ characteristic that displays no hysteresis, but then have no pull-in instability either, i.e., $H_V = 0$ & $y_c = 0$, as shown in Fig. 5A. Since $H_V \propto y_c^2$ (see section A1 of Appendix for derivation) and $y_c$ given by Eq. 5, $y_d^{eff} = 2y_0$ makes both $H_V$ and $y_c$ zero. It means that a series capacitor $C_s = \epsilon_0 / y_d^{eff}$ (semiconductor depletion capacitance) with half the value of the initial air-gap capacitance ($\epsilon_0 / y_0$) in Fig. 1B, enables hysteresis-free smooth switching. Note that, this switching behavior has fundamentally been made possible by stabilizing the gate in its inherently unstable regime, as shown in Fig. 3D. This stabilization comes from an inherent negative feedback provided by the series capacitor ($C_s$), so that the voltage-drop across air-gap capacitor ($V_{air}$) decreases when the gate enters in the unstable regime (Fig. 5B). This decrease in $V_{air}$ amplifies the voltage-drop across $C_s$ (channel potential ($\psi_s$)) (Fig. 5B). This amplification in $\psi_s$ is directly reflected in the body factor $m < 1$, symbols in Fig. 5C. If one accounts for the charge build up inside the semiconductor (i.e., voltage dependence of $y_d^{eff}$), it can be shown that (see section A3 of Appendix for derivation)-

$$m \equiv \frac{dV_G}{d\psi_s} = 1 + \frac{C_s}{C_{air}^{eff}}, \qquad 7(a)$$

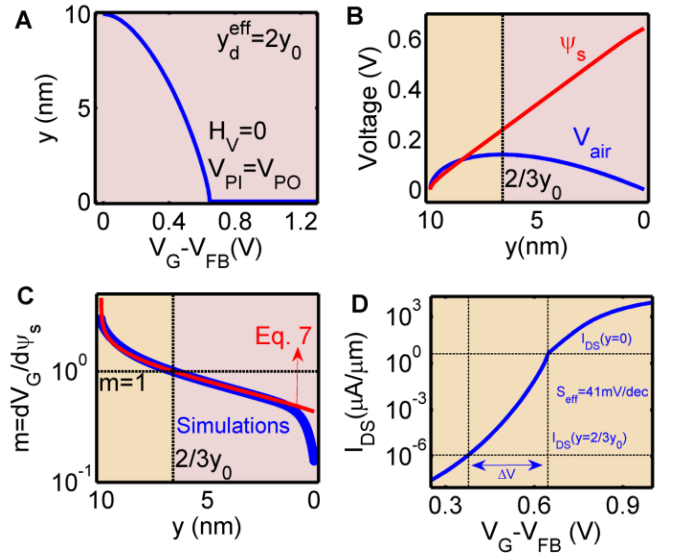

**Fig. 5:** Hysteresis-free sub-$60 mV/dec$ switching in NEMFET. **(A)** $y - V_G$ characteristic for hysteresis-free smooth switching. **(B)** Voltage-drop across air-gap capacitor ($V_{air}$) and series capacitor ($\psi_s$) showing the voltage amplification in $\psi_s$ in the unstable regime. **(C)** Body factor $m$ is less than one in the unstable regime of NEMFET. Symbols denote the numerical simulations and solid line Eq. 7. **(D)** Corresponding $I_{DS} - V_G$ characteristic with an effective sub-threshold swing of $41 mV/dec$ reflecting the voltage amplification provided by the negative capacitor. The parameters used in the simulations are $y_0 = 10 nm$, $y_d = 0.5 nm$, $\epsilon_s = 11.7$, $\epsilon_r = 3.9$, $L = 20 nm$, $W = 4 \mu m$, $H = 26.4 nm$, $E = 200 GPa$, $N_A = 4.95e15 cm^{-3}$, $V_{DS} = 0.5V$.

$$C_s = \sqrt{\frac{q \epsilon_0 \epsilon_s N_A}{2\psi_s}}, \quad C_{air}^{eff} \approx \frac{\epsilon_0}{3\left(y - \frac{2}{3}y_0\right)}, \qquad 7(b)$$

where $q$ is the charge on an electron, $\epsilon_s$ is dielectric constant of channel material, and $N_A$ is the channel doping. $C_{air}^{eff}$ is the effective air-gap capacitor that determines the coupling between $V_G$ and $\psi_s$. *Interestingly, air-gap capacitor acts effectively like a negative capacitor when gate enters in the unstable regime (i.e., $C_{air}^{eff} < 0$ when $y < \frac{2}{3}y_0$, Eq. 7(b)) and therefore, provides necessary voltage amplification to reduce $m$ below one.* Equation 7 correctly reproduces the numerical simulations results (see for SI-II for numerical simulations framework and Ref. [15]) in Fig. 5C.

It is important to note that the value of $N_A$ must be carefully optimized so that it provides the necessary series capacitance to stabilize the gate throughout the air-gap. The corresponding $I_{DS} - V_G$ for the same NEMFET obtained from the self-consistent numerical simulations (see section A2 of appendix for numerical simulations framework) is shown in Fig. 5D which indeed confirms the hysteresis-free smooth switching of NEMFET. In sub-threshold regime, $I_{DS} - V_G$ characteristic is highly nonlinear and does not exhibit a constant sub-threshold swing. Therefore, we define an effective sub-threshold swing ($S_{eff}$) for $I_{DS}\left(y = \frac{2}{3}y_0\right) < I_{DS} < I_{DS}(y = 0)$ when NEMFET is biased in negative capacitance regime. Considering this, $S_{eff}$ is given by $\Delta V / \log\left(\frac{I_{DS}(y=0)}{I_{DS}(y=\frac{2}{3}y_0)}\right)$, where $\Delta V$ is defined in Fig. 5D. The value of $S_{eff}$ for the chosen parameters is capacitance regime.

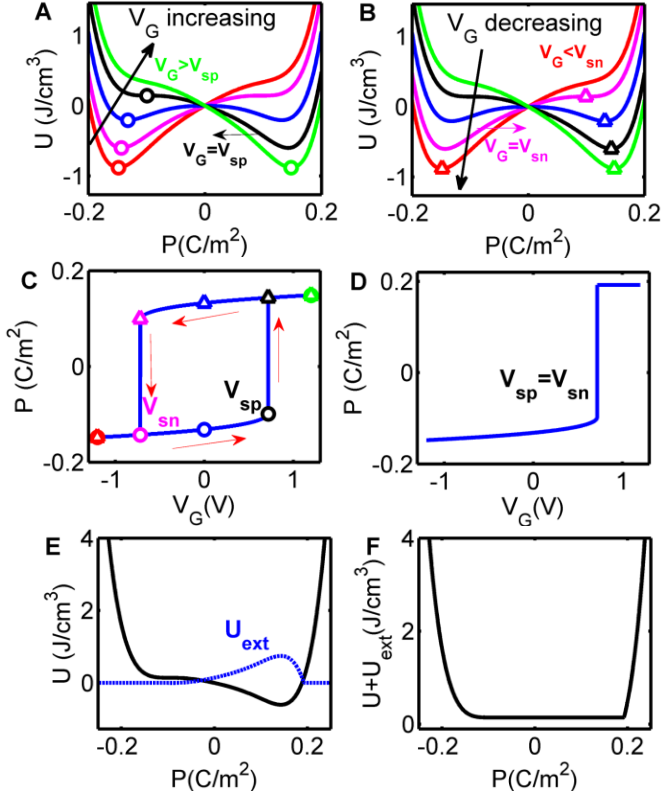

**Fig. 6:** Prospects of Hysteresis-free abrupt switching in FE-FET-another example of a Landau switch. Energy landscapes when **(A)** $V_G$ is increasing and **(B)** $V_G$ is decreasing. Open circle and open triangle denotes the equilibrium position. **(C)** The corresponding $P - V_G$ characteristic of ferroelectric insulator showing the abrupt jump in polarization responsible for $S = 0mV/dec$ switching. **(D)** $P - V_G$ characteristic required for hysteresis-free abrupt switching and **(E)** extra energy component ($U_{ext}$) needed to compensate the energy barrier present at $V_G = V_{sp}$ to make **(F)** energy landscape flat. Parameters ($\alpha = -10^7 m/F, \beta = -8.9 \times 10^8 m^5/F/C^2, \gamma = 4.5 \times 10^{10} m^9/F/C^4$) [16] used for simulation are of a typical ferroelectric dielectric material BaTiO$_3$.

$41mV/dec$ and is less than the fundamental thermodynamiclimit of $60mV/dec$. Reduction in $S_{eff}$ confirms the voltage amplification provided by the air-gap capacitor in its negative

Note that, the idea of using a series capacitor to achieve $y_c = 0$ is long known in MEMS literature [17], [18], however, its implications as a negative capacitor for voltage amplification in a FET has not been appreciated. Finally, the hysteresis-free sub$-60mV/dec$ switching in NEMFET is obtained by utilizing the nonlinear electromechanical coupling between electrostatic and mechanical energy/forces of NEMFET. As opposed to other sub$-60mV/dec$ switching schemes, such as Ferroelectric-FET with negative capacitance [16], Tunnel-FETs [4], and Impact Ionization FETs [5], NEMFET is unique as it only utilizes the electromechanical coupling, rather than any specific material property, to achieve hysteresis-free sub-60-$mV/dec$ switching.

## V. DISCUSSIONS ON FE-FET

Based on the analysis of NEMFET, let us summarize the general conclusions regarding Landau switches-

- Hysteresis-free abrupt switching ($S = 0$) in Landau switches require an extra energy component to go from an intrinsic two well energy landscape to a flat energy landscape.
- Hysteresis-free sub$-60mV/dec$ switching is obtained by stabilizing the unstable gate insulator in its unstable regime.

We now validate the generality of these conclusions using FE-FET-another example of a Landau switch. We again follow the same procedure and look at the evolution of energy landscapes of FE-FET. Total energy ($U$) of a ferroelectric dielectric system in terms of its polarization ($P$) is given by-

$$U = \alpha P^2 + \beta P^4 + \gamma P^6 - PE, \qquad (8)$$

where $\alpha, \beta$ and $\gamma$ are material dependent constants and $E = V_G/y$ is the applied electric field. Figure 6(A) shows energy landscapes when $V_G$ is increasing assuming that dielectric is negatively polarized to begin with ($P < 0$ at $V_G = 0$). Open circles denote the position of stable equilibrium. As $V_G$ increases, energy landscape changes such that value of $P$ at equilibrium increases (though keeping the same negative sign, see Fig. 6C). Beyond a certain $V_G > V_{sp}$, however, the left minima of energy landscape (occurring at $P < 0$, Fig. 6A) vanishes and $P$ has to abruptly switch from a negative value to a positive value (Fig. 6A, energy landscape corresponding to $V_G > V_{sp}$). Like a NEMFET, if $V_G$ is reduced below $V_{sp}$, $P$ can not switch back from a positive value to a negative value, because of the presence of an energy barrier (Figs. 6A-B). $V_G$ has to be reduced below $V_{sn}$ for switching the polarization back to a positive value (see Figs. 6B-C) and that causes the hysteretic $P - V_G$ characteristic as shown in Fig. 6C.

For hysteresis-free abrupt switching in FE-FET, a $P - V_G$ characteristic as shown in Fig. 6D is required (similar to the $y - V_G$ characteristic of NEMFET in Fig. 4A). Similar to a NEMFET, such $P - V_G$ characteristic can only be achieved if energy barrier at $V_G = V_{sp}$ can be compensated by an extra energy component ($U_{ext}$) as shown in Fig. 6E to give rise to a flat energy landscape in Fig. 6F. *Therefore, the requirement of an extra energy component to make energy landscape flat is the requirement of all landau switches.*

Regarding hysteresis-free smooth switching in FE-FET, it has previously been shown that stabilizing the ferroelectric in its unstable regime using a series capacitor (like NEMFET in Fig. 5) puts it into negative capacitance regime and exhibits sub-60-$mV/dec$ switching [16], [19], [20]. *Although the detailed physics of NEMFET and FE-FET is completely different, the general conclusions regarding hysteresis-free abrupt and smooth switching are essentially the same.*

## VI. CONCLUSIONS

We have shown that FE-FET and NEMFET belong to a general class of switches called Landau switches characterized by a two-well energy landscape similar to that of a phase transition. We determined that an extra energy component is needed to make the energy landscape of Landau switches flat to achieve hysteresis free abrupt switching. In contrast, stabilization of the unstable gate insulator in its unstable regime provides necessary internal voltage amplification to achieve hysteresis-free sub-$60mV/dec$ switching. NEMFET provides a unique material independent way to achieve hysteresis-free sub-$60mV/dec$ only by utilizing nonlinear electromechanical coupling unlike any other proposal of sub-$60mV/dec$ switch e.g., Tunnel FETs, Impact Ionization FETs, Ferroelectric negative capacitance FETs. Although, the results are discussed for top of the barrier transport, the general conclusions are valid for other transport mechanisms also.

### ACKNOWLEDGEMENT

We gratefully acknowledge discussions with Prof. Mark Lundstrom (Purdue University), financial support from MSD-FCRP center, PRISM center and computational resources from Network for Computational Nanotechnology (NCN).

### APPENDIX

*A1-Pull-in instability and hysteresis*

The static behavior of NEMFET is governed by the minimization of total system energy $U = U_s + U_e$ (Eqs. 3-4 in main text). Minimization of $U$ with respect to the gap ($y$) i.e. $\frac{dU}{dy} = 0$, yields the following force balance equation-

$$k(y_0 - y) = \frac{1}{2} \frac{\epsilon_0 A}{\left(y_d^{eff} + y\right)^2} V_G^2, \qquad (A1)$$

where left hand side of Eq. A1 is the spring force and right hand side is the electrostatic force. Pull-in instability occurs when $U - y$ profile exhibits an inflection point i.e. $\frac{d^2 U}{dy^2} = 0$, that yields-

$$k = \frac{\epsilon_0 A}{\left(y_d^{eff} + y\right)^3} V_G^2. \qquad (A2)$$

Solution of Eqs. A1-A2 gives the critical gap $y_c$ at which pull-in instability occurs-

$$y_c = \frac{2}{3} y_0 - \frac{1}{3} y_d^{eff}, \qquad (A3)$$

which is same as Eq. 3 in the main text. Using Eq. A1 & A3 value of pull-in voltage ($V_{PI}$) is given by-

$$V_{PI} = \sqrt{\frac{8}{27} k \frac{\left(y_0 + y_d^{eff}\right)^3}{\epsilon_0 A}}. \qquad (A4)$$

Once the gate is pulled-in, it does not spring back at the same voltage as discussed in the main text, rather voltage has to be reduced below pull-out voltage ($V_{PO}$). Expression of $V_{PO}$ can simply be obtained by putting $y = 0$ in Eq. A1 and is given by-

$$V_{PO} = \sqrt{2k \frac{y_0 \left(y_d^{eff}\right)^2}{\epsilon_0 A}}. \qquad (A5)$$

Now using the analytical formula of $V_{PI}$ and $V_{PO}$, expression for hysteresis $H_V = V_{PI} - V_{PO}$ is given by-

$$H_V = \frac{8k}{3} \frac{\left(4y_d^{eff} + y_0\right)}{V_{PI} + V_{PO}} y_c^2, \qquad (A6)$$

which suggests that $y_c = 0$ implies $H_V = 0$.

*A2-Simulation framework for NEMFET*

In a practical NEMFET the buildup of charge inside the channel should be taken into account. Considering that, the static behavior of NEMFET is dictated by the balance of spring and electrostatic forces, i.e.

$$k(y_0 - y) = \frac{1}{2} \epsilon_0 E_{air}^2 A, \qquad (A7)$$

where $E_{air}$ is the electric field in the air and is equal to $\epsilon_s E_s(\psi_s)$, where, $\epsilon_s$ is the dielectric constant of the substrate, and

$$E_s(\psi_s) = \sqrt{\frac{2qN_A}{\epsilon_0 \epsilon_s}} \left[ \psi_s + \left( e^{-\frac{q\psi_s}{k_B T}} - 1 \right) \frac{k_B T}{q} \right.$$
$$\left. - \left( \frac{n_i}{N_A} \right)^2 \left( \psi_s - \left( e^{\frac{q\psi_s}{k_B T}} - 1 \right) \frac{k_B T}{q} \right) \right]^{\frac{1}{2}}, \quad (A8)$$

where, $E_s(\psi_s)$ is the electric field at the substrate-dielectric interface, $\psi_s$ is the surface potential, $q$ is the charge on an electron, $N_A$ is the substrate doping, $k_B$ is the Boltzmann constant, $T$ is the absolute temperature, and $n_i$ is the intrinsic carrier concentration in the substrate. Voltage drop in air ($y \epsilon_s E_s(\psi_s)$), dielectric $\left( \frac{y_d}{\epsilon_d} \epsilon_s E_s(\psi_s) \right)$, and substrate($\psi_s$) can be related to the applied gate bias $V_G$ as follows:

$$V_G = V_{FB} + \left( y + \frac{y_d}{\epsilon_d} \right) \epsilon_s E_s(\psi_s) + \psi_s, \qquad (A9)$$

where, $y_d$ is the dielectric thickness and $V_{FB}$ is the flat band voltage. Equations A7-A9 are solved self-consistently for $y$ and $\psi_s$ at each $V_G$. The corresponding inversion charge density ($Q_i$) in the channel and drain current ($I_{DS}$) are given by,

$$Q_i = \frac{qn_i^2}{N_A} \int_0^{\psi_s} \frac{e^{\frac{q\psi}{k_B T}}}{E_s(\psi)} d\psi, \qquad (A10)$$

$$I_{DS} = \mu_n L Q_i \frac{V_{DS}}{W}, \qquad (A11)$$

where, $\mu_n$ is the channel mobility for electrons, $V_{DS}$ is the applied drain to source voltage.

*A3- Derivation of body factor ($m$) for NEMFET*

In order to derive body factor $m = dV_G/d\psi_s$, we consider the sub-threshold regime in which electrostatic force is given by $q\epsilon_s N_A A\psi_s$ (using Eqs. A7-A8). Therefore, equation A7 reduces to-

$$k(y_0 - y) = q\epsilon_s N_A A\psi_s. \qquad (A12)$$

Similarly, Eq. A9 reduces to-

$$V_G = V_{FB} + \psi_s + \left(y + \frac{y_d}{\epsilon_d}\right)\beta\sqrt{\psi_s}, \qquad (A13)$$

where $\beta$ is $\sqrt{\frac{2q\epsilon_s N_A}{\epsilon_0}}$. Now using $y = y_0 - \left(\frac{q\epsilon_s N_A A}{k}\right)\psi_s$ (from Eq. A12), Eq. A13 reduces to-

$$V_G = V_{FB} + \psi_s + \left(y_0 - \left(\frac{q\epsilon_s N_A A}{k}\right)\psi_s + \frac{y_d}{\epsilon_d}\right)\beta\sqrt{\psi_s}. \qquad (A14)$$

Now taking derivative of Eq. A14 with respect to $\psi_s$, we get-

$$\frac{dV_G}{d\psi_s} = 1 + \left(y_0 + \frac{y_d}{\epsilon_d}\right)\frac{\beta}{2\sqrt{\psi_s}} - \frac{3}{2}\left(\frac{q\epsilon_s N_A A}{k}\right)\sqrt{\psi_s}, \qquad (A15)$$

Simplifying this we get-

$$m \equiv \frac{dV_G}{d\psi_s} = 1 + \frac{C_s}{C_{air}^{eff}}, \qquad (A16)$$

$$C_{air}^{eff} = \frac{\epsilon_0}{3\left(y - \frac{2}{3}y_0 + \frac{y_d}{3\epsilon_r}\right)} \approx \frac{\epsilon_0}{3\left(y - \frac{2}{3}y_0\right)}. \qquad (A17)$$

where $C_s$ is the depletion capacitance and $C_{air}^{eff}$ is the effective air-gap capacitance. Equations A16-A17 predicts that air-gap capacitor acts as a negative capacitor if gate is stabilized in its unstable regime i.e. $y < \frac{2}{3}y_0$ and thus making $m < 1$.

## REFERENCES

[1] T. N. Theis and P. M. Solomon, "In Quest of the Next Switch: Prospects for Greatly Reduced Power Dissipation in a Successor to the Silicon Field-Effect Transistor," *Proceedings of the IEEE*, vol. 98, no. 12, 2010.

[2] V. V. Zhirnov and R. K. Cavin, "Negative capacitance to the rescue ?," *Nature Nanotechnology*, vol. 3, pp. 77-78, Feb. 2008.

[3] T. Yuan and N. T. H., *Fundamentals of Modern VLSI Devices*. Cambridge University Press.

[4] A. M. Ionescu and H. Riel, "Tunnel field-effect transistors as energy-efficient electronic switches.," *Nature*, vol. 479, no. 7373, pp. 329-37, Nov. 2011.

[5] P. B. G. Kailash Gopalakrishnan and J. D. Plummer, "I-MOS: A Novel Semiconductor Device with a Subthreshold Slope lower than kT/q," in *IEDM*, 2002.

[6] J.-C. Tolédano and P. Tolédano, *The Landau theory of phase transitions*. Singapore: World Scientific, 1987.

[7] W. Shu-Yau, "A New Ferroelectric Memory Device, Metal-Ferroelectric-Semiconductor Transistor," *Transactions on Electron Devices*, vol. 201, no. 8, pp. 499-504, 1974.

[8] H. Kam, D. T. Lee, R. T. Howe, and T.-J. King, "A new nano-electro-mechanical field effect transistor (NEMFET) design for low-power electronics," in *IEDM*, 2005, pp. 463-466.

[9] H. C. Nathanson, W. E. Newell, R. A. Wickstrom, and J. R. Davis, "The Resonant Gate Transistor," *Transactions on Electron Devices*, vol. 14, no. 3, pp. 117-133, 1967.

[10] N. Abele, R. Fritschi, K. Boucart, F. Casset, P. Ancey, and A. M. Ionescu, "Suspended-Gate MOSFET: bringing new MEMS functionality into solid-state MOS transistor," in *IEDM*, 2005, vol. 0, no. c.

[11] S. Krylov, "Lyapunov exponents as a criterion for the dynamic pull-in instability of electrostatically actuated microstructures," *International Journal of Non-Linear Mechanics*, vol. 42, pp. 626-642, 2007.

[12] V. M. B. David M. Burns, "Nonlinear flexures from stable deflection of an electrostatically actuated micromirror," *SPIE*, vol. 3226, pp. 125-136, 1997.

[13] E. M. Abdel-rahman, M. I. Younis, and A. H. Nayfeh, "Characterization of the mechanical behavior of an electrically actuated microbeam," *Journal of Micromechanics and Microengineering*, vol. 12, p. 759, 2002.

[14] E. K. Chan, E. C. Kan, and R. W. Dutton, "Nonlinear Dynamic Modeling of Micromachined Switches," *IEEE MTT-S Digest*, vol. 3, pp. 1511-1514, 1997.

[15] A. Jain, P. R. Nair, and M. A. Alam, "Flexure-FET biosensor to break the fundamental sensitivity limits of nanobiosensors using nonlinear electromechanical coupling.," *Proceedings of the National Academy of Sciences of the United States of America*, vol. 109, no. 24, pp. 9304-8, Jun. 2012.

[16] S. Salahuddin and S. Datta, "Use of negative capacitance to provide voltage amplification for low power nanoscale devices.," *Nano letters*, vol. 8, no. 2, pp. 405-10, Feb. 2008.

[17] J. I. Seeger and S. B. Crary, "Stabilization of electrostatically actuated mechanical devices," in *Transducers*, 1997, pp. 1133-1136.

[18] E. K. Chan and R. W. Dutton, "Electrostatic Micromechanical Actuator with Extended Range of Travel," *Journal of Microelectromechanical Systems*, vol. 9, no. 3, pp. 321-328, 2000.

[19] S. Salahuddin and S. Datta, "Can the subthreshold swing in a classical FET be lowered below 60 mV/decade?," in *IEDM*, 2008.

[20] A. I. Khan, C. W. Yeung, C. Hu, and S. Salahuddin, "Ferroelectric Negative Capacitance MOSFET : Capacitance Tuning & Antiferroelectric Operation," in *IEDM*, 2011, pp. 255-258.

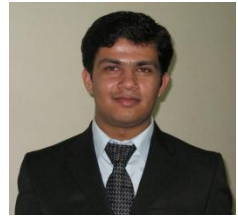

**Ankit Jain** received the B.Tech. degree in electrical engineering from the Indian Institute of Technology Kanpur, Kanpur, India, in May 2008. He is currently working toward the Ph.D. degree in the school of Electrical and Computer Engineering, Purdue University, West Lafayette, IN, under the guidance of Prof. M. A. Alam.

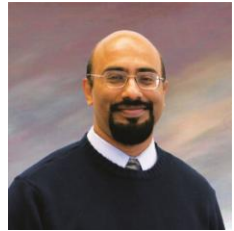

**Muhammad Ashraful Alam** (M'96–SM'01–F'06) received the Doctoral degree from Purdue University, West Lafayette, IN, in 1995. He is a Professor of electrical and computer engineering and a University Faculty Scholar at Purdue University, where his research and teaching focus on physics, simulation, characterization, and technology of classical and emerging electronic devices.